\documentclass[a4paper, twocolumn]{article}

\usepackage[english]{babel}
\usepackage{graphicx}
\usepackage{xcolor}
\usepackage{amsmath}
\usepackage{amsfonts}
\usepackage{authblk}
\usepackage[inline]{enumitem}
\usepackage[normalem]{ulem}

\usepackage[a4paper, margin=0.7in]{geometry}
\usepackage{setspace}

\usepackage{hyperref}
\hypersetup{          
 pdftitle={Observing a Scale Anomaly in Graphene: A Universal Quantum Phase Transition},
 pdfauthor={Omrie Ovdat, Jinhai Mao, Yuhang Jiang, Eva Y. Andrei, and Eric Akkermans},
 pdfsubject={},
 pdfkeywords={},
 unicode=true,
 breaklinks=true,
 linkcolor=blue, 
 citecolor=[rgb]{1,0.0,0.0},
 pdfborder={0 0 0},
 pdfborderstyle={},
 colorlinks=true,
 bookmarksnumbered=true,
 bookmarksopen=true,
 bookmarksopenlevel=1,
}

\usepackage{pdfpages}

\usepackage[sort&compress,numbers]{natbib}
\def\be{\begin{equation}}
\def\ee{\end{equation}}
\newcommand{\email}[1]{\href{mailto:#1}{#1}}
\makeatletter
\def\blfootnote{\xdef\@thefnmark{}\@footnotetext}
\makeatother

\begin{document}

\title{{\bf Observing a scale anomaly and a universal
quantum phase transition in graphene}}

\author[1,*]{O. Ovdat}
\author[2,*]{Jinhai Mao}
\author[2]{Yuhang Jiang}
\author[2]{E. Y. Andrei}
\author[1]{E. Akkermans}
%
\affil[1]{Department of Physics, Technion -- Israel Institute of Technology, Haifa 3200003, Israel}
\affil[2]{Department of Physics and Astronomy, Rutgers University,\par{}Piscataway, New Jersey 08854}

\date{}
%
%
\maketitle

\begin{abstract}
 
One of the most interesting predictions resulting from quantum physics, is the violation of classical symmetries, collectively referred to as anomalies. A remarkable class of anomalies occurs when the continuous scale symmetry  of a scale free quantum system is broken into a discrete scale symmetry for a critical value of a control parameter. This is an example of a (zero temperature) quantum phase transition. Such an anomaly takes place for the  quantum  inverse square potential known to describe  'Efimov physics'. Broken continuous scale symmetry into discrete scale symmetry also appears for a charged and massless Dirac fermion in an attractive $1/r$ Coulomb potential. The purpose of this article is to demonstrate the universality of this  quantum phase transition and to present convincing experimental evidence of its existence for a charged and massless fermion in an attractive Coulomb potential as realised in graphene.
\blfootnote{\llap{$^*~$}These authors contributed equally to this work.}
\blfootnote{Correspondence should be addressed to E.~Akkermans (\email{eric@physics.technion.ac.il})}
\end{abstract}

\begin{figure*}[!t]
\centering
\includegraphics[width= \textwidth, height = \textheight, keepaspectratio]{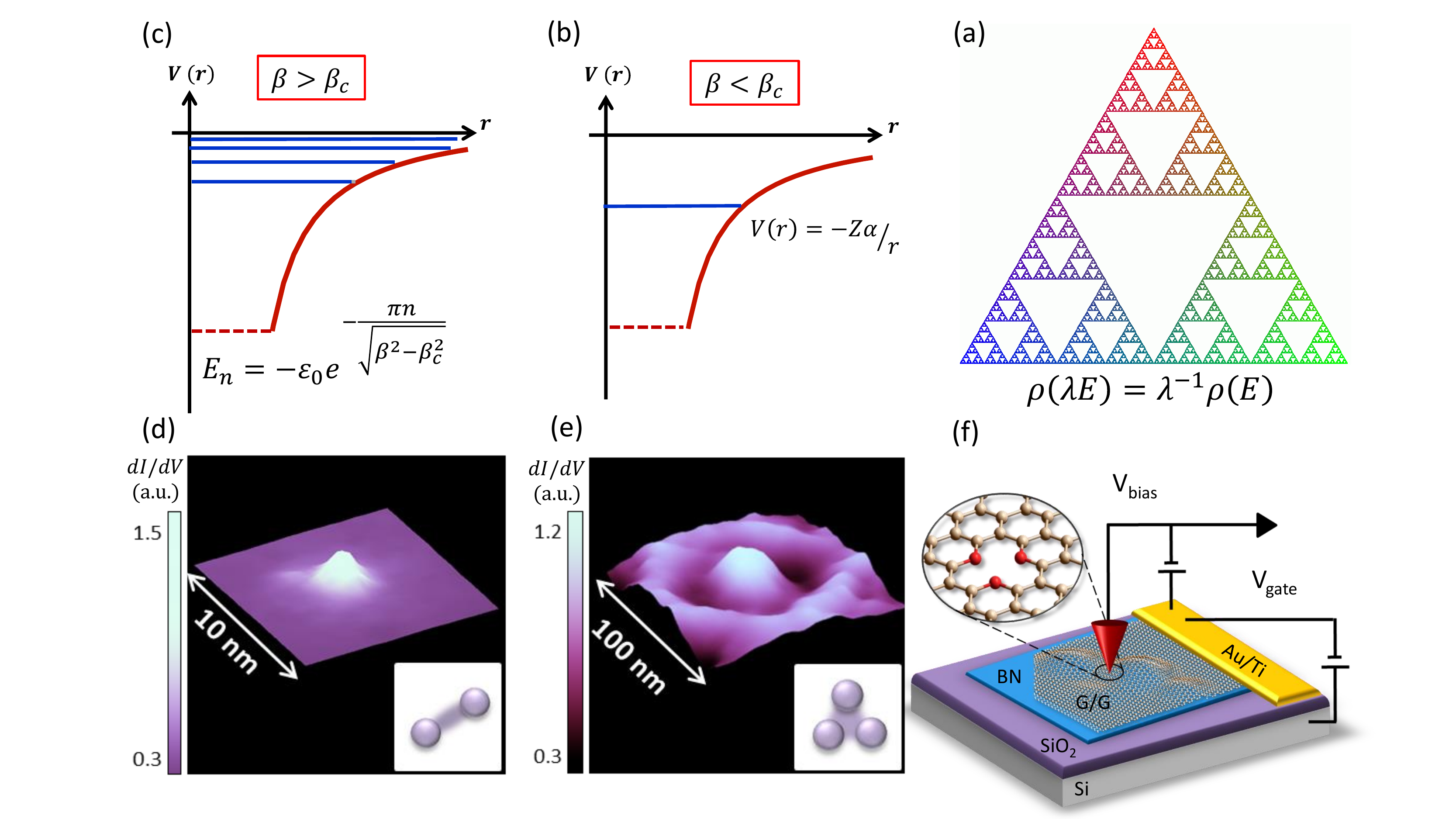}
\protect\caption{\label{fig1}
\textbf{Schematic visualization of the purpose and main results of this paper}. (a) Sierpinski gasket as typical featuring of such iterative fractal structures. This QPT is realized experimentally by creating single-atom vacancies in graphene. The function $\rho(E)$ is the density of states and obeys a scaling relation characterising the existence of discrete scale symmetry. (b)-(c) Illustration of the universal Quantum Phase Transition (QPT) obtained by varying  the dimensionless parameter $\beta \equiv Z\alpha$ (see text for precise definitions) in the low energy spectrum of a massless fermion in a Coulomb potential $V = -Z\alpha/r$ created by a charge $Z$. (b) For low values $\beta<\beta_{\mathrm{c}}$, there is a single quasi-bound state close to zero energy. (c) For overcritical values $\beta>\beta_{\mathrm{c}}$, the low energy spectrum is a ladder $E_n$ characterized by a discrete scale symmetry $\{E_n \} = \{\lambda E_n \}$ for $\lambda = \mathrm{exp}(\pi/\sqrt{\beta^2-\beta_{\mathrm{c}}^2})$. (d)-(e) Experimental $dI/dV$ maps of charged vacancy for fixed $\beta<\beta_{\mathrm{c}}$ (d) and $\beta>\beta_{\mathrm{c}}$ (e). The images illustrate the characteristic probability density of the resonances in (b) and (c). (f) Scanning tunnelling microscopy (STM) setup. Local charge $Z$ is accumulated at the single vacancy in graphene by applying voltage pulses to the STM tip.}
\end{figure*}

\vspace{5mm}

Continuous scale symmetry (CS) -- a common property of physical systems -- expresses the invariance of a physical quantity $f (x)$  (e.g., the mass) when changing a control parameter $x$ (e.g., the length). This property is expressed by a simple scaling relation, $f(ax) = b \, f(x)$, satisfied  $\forall a>0$ and corresponding $b(a)$, whose  general solution is the power law $f(x) = C \, x^\alpha$ with $\alpha = \ln b / \ln a$. Other physical systems possess the weaker discrete scale symmetry (DS) expressed by the same aforementioned scaling relation but now satisfied for fixed values $(a,b)$ and whose solution becomes $f(x) = x^\alpha \, G\left( \ln x / \ln a \right)$, where $G(u+1) = G(u)$ is a periodic function.  Physical systems having a DS are also known as self-similar fractals~\cite{akkermans2013statistical} (Fig.~\ref{fig1}a). It is possible to break CS into DS at the quantum level, a result which constitutes the basis of a special kind of scale anomaly \cite{Adler69,jackiw69}.

A well studied example is provided by the problem of a particle of mass $\mu$ in an attractive inverse square potential \cite{case1950, landau1991quantum} which plays a role in various systems~\cite{levy1967electron,CamblongEpeleFanchiottiEtAl2001,Kaplan:2009kr,nisoli2014attractive} and more importantly in Efimov physics~\cite{efimov1970energy,Efimov1971}. Although well defined classically, the quantum mechanics of the scale -- and conformal~\cite{jackiw1995diverse} -- invariant Hamiltonian $H = -  \Delta / 2 \mu  - \xi / r^2$   (with $\hbar = 1$) is well posed, but for large enough values of $\xi$, $H$ is no longer self-adjoint~\cite{meetz1964singular,gitman2012self}. The corresponding Schr\"odinger equation for a normalisable wave function $\psi (r)$ of energy $k^2 = - 2 \mu E$ is, 
\be 
\psi '' (r) + \frac{d -1}{r} \psi' (r) + \frac{\zeta}{r^2} \psi (r) = k^2 \psi(r)
\label{eq1}
\ee
where $\zeta \equiv 2 \mu \xi - l ( l+d-2)$ is a dimensionless parameter, $d$ the space dimensionality and $l$ the orbital angular momentum. Equation (\ref{eq1}) is invariant under the transformation $r \rightarrow \lambda \, r $ and $k \rightarrow k / \lambda$, $\forall \lambda$ (CS), namely to every normalisable wave function of energy $k^2$ corresponds a continuous family of states with energies $(\lambda k)^2$, so that the bound spectrum is a continuum unbounded from below. Various ways exist to cure this problem,  based on cutoff regularisation and 
 renormalisation group~\cite{albeverio1981class, Beane:2000wh, mueller2004renormalization, Braaten2004, HammerSwingle2006a, MorozSchmidt2009, PhysRevLett.112.186603}, and all lead for the low energy spectrum to a  quantum phase transition (QPT) monitored by $\zeta$, between a single bound state for $\zeta < \zeta_{\mathrm{c}}$ to an infinite and discrete energy spectrum for $ \zeta > \zeta_{\mathrm{c}}$, independent of the regularisation procedure and given by  
\be
 k_n(\zeta) = \epsilon_0 \, e^{- \frac{ \pi  n }{ \sqrt{\zeta - \zeta_{\mathrm{c}}}}}, \, \, \, n \in \mathbb{Z} 
\label{eq2}
\ee
 which clearly displays DS. The critical value $\zeta_{\mathrm{c}} = (d-2)^2 / 4$ depends on the space dimensionality only, and $\epsilon_0$ is a regularization dependent energy scale. In the overcritical phase $\zeta > \zeta_{\mathrm{c}}$, the corresponding renormalization group solution provides a rare example of a limit cycle~\cite{PhysRevB.46.12664, albeverio1981class, Beane:2000wh}. Building on the previous example, it can be anticipated that the problem of a massless Dirac fermion in an attractive Coulomb potential \cite{shytov2007atomic,PhysRevLett.99.166802,PhysRevB.90.165414}), $- Z \alpha / r$, is also scale invariant (CS) and that the spectrum of resonant quasi-bound states presents similar features and a corresponding QPT. 
 
In this work, we demonstrate the existence of such a universal QPT for arbitrary space dimension $d \geq 2$ and independently of the short distance regularisation. We obtain an explicit formula for the low energy fractal spectrum in the overcritical regime. In contrast to the Schr\"odinger case \eqref{eq1}, the massless Dirac Hamiltonian displays an additional parity symmetry which may be broken by  the regularisation. In that case, the degeneracy of the overcritical fractal spectrum is removed and two intertwined geometric ladders of quasi bound states appear in the $s$-wave channel.  All these features are experimentally demonstrated using a charged vacancy in graphene. We observe the overcritical spectrum and we obtain an experimental value for the universal geometric ladder factor in full agreement with the theoretical prediction. We also explain the observation of two intertwined ladders of quasi bound states as resulting from the breaking of parity symmetry.  Finally, we relate our findings to Efimov physics as measured in cold atomic gases.

\subsection*{Results}
\textbf{The Dirac model.} The Dirac equation of a massless fermion in the presence of a $-Z\alpha/r$ potential is obtained from the Hamiltonian (with $\hbar = c = 1$),
\be
H = -i \gamma^0 \gamma^j \partial_j - \frac{\beta }{ r} 
\label{dirac} \ee
where $( \gamma^0 , \gamma^j )$ are Dirac matrices. Here the dimensionless parameter monitoring the transition is $\beta = Z \alpha$, where $Z$ is the Coulomb charge and $\alpha$ the fine structure constant. The QPT occurs at the critical value $\beta_{\mathrm{c}} = (d-1)/2$ \footnote{A related anomalous behaviour in the Dirac Coulomb problem has been identified long ago~\cite{pomeranchuk1945energy} but its physical relevance was marginal since it required non existent heavy-nuclei Coulomb charges $Z \simeq 1 / \alpha \simeq 137$ to be observed. Moreover, the problem of a massive Dirac particle is different due to the existence of a finite gap which breaks CS.} (see Supplementary Note 1). For resonant quasi-bound states, we look for scattering solutions of the form $\psi_{\mathrm{in}} + e^{2i \eta} \psi_{\mathrm{sc}}$, where $\eta (E)$ is the energy-dependent scattering phase shift and $\psi_{\mathrm{in,sc}} (r,E)$ are two component objects representing the radial part of the 
 Dirac spinor which behave asymptotically as,
\be 
\psi_{\mathrm{in,sc}} (r, E) = r^{ \frac{1 - d }{2}} \left( V_{\mathrm{in,sc}} \, \,  (2 i |E| r)^{\mp i \beta} \, e^{\mp iEr} \right)
\label{near origin} 
\ee 
for $ |E| r \gg 1 $ and, using $\gamma \equiv \sqrt{\beta^2 - \beta_{\mathrm{c}}^2}$,
\be
\psi_{\mathrm{in,sc}} (r, E) = r^{\frac{1 - d }{ 2}} \left( U_{\mathrm{in,sc}} ^- \, \,  (2 i E r)^{ - i\gamma} + U_{\mathrm{in,sc}} ^+ \, \,  (2 i E r)^{ i\gamma} \right),  
\label{asympt} 
\ee
for $|E| r \ll 1 $ and for the lowest angular momentum channels. The two component objects $V_{\mathrm{in,sc}}$ and $U_{\mathrm{in,sc}}^{\pm}$ in equations \eqref{near origin},\eqref{asympt} are constants. It is easy to infer from (\ref{asympt}) that $\beta = \beta_{\mathrm{c}}$ plays a special role. Indeed for $\beta > \beta_{\mathrm{c}}$, there exists a family of normalisable solutions which admit complex eigenvalues $E = - i \epsilon$, hence the Hamiltonian (\ref{dirac}) is not self-adjoint $( H \neq H^\dagger )$. To properly define this quantum problem, a regularisation is thus needed for the too strong potential at overcritical values of $\beta = Z \alpha$. This is achieved by introducing a cutoff length $L$ and a boundary condition at $r = L$, which is equivalent to replacing the Coulomb potential at short distances by a well behaved potential whose exact form is irrelevant in the low energy regime $E L \ll 1$. The resulting mixed boundary condition can be written as $h = \left.\Psi_2\left(r,E\right)/\Psi_1\left(r,E\right)\right\vert_{r\rightarrow L^+}$, where $\Psi_{1,2}$ represent the two components of the aforementioned radial part of the Dirac spinor. The resulting scattering phase shift $\eta \left(E, L,h \right) $, which contains all the information about the regularisation, thus becomes a function of $L$ and of the parameter $h$. The quasi-bound states energy spectrum is obtained from the scattering phase shift by means of the  Krein-Schwinger relation~\cite{EA, ADL} which relates the change of density of states $\delta \rho$ to the energy derivative of $\eta$,\footnote{This is also related to the Wigner time delay~\cite{smith} and to the Friedel sum rule}
\be 
\delta \rho (E ) = \frac{1 }{ \pi } \frac{ d \eta (E ) }{ d E} \, .
\label{ks} \ee

\textbf{Theoretical structure of quasi-bound spectrum}. From now on, and to compare to experimental results further discussed, we consider the case $d = 2$ for which there is a single orbital angular momentum quantum number $m \in \mathbb{Z}$. The corresponding critical coupling becomes $\beta_{\mathrm{c}} = |m+1/2| \geq 1/2$, giving rise to the $s$-wave channels, $m = 0,-1$ for which $\beta_{\mathrm{c}} = 1/2$. Depending on the choice of boundary condition $h$, $\delta \rho (E )$ can be degenerate or non-degenerate over these two $s$-wave channels. This degeneracy originates from the symmetry of the $(2 + 1)$ Dirac Hamiltonian \eqref{dirac} under parity, $(x,y)\rightarrow (-x,y)$, and its existence is equivalent to whether or not the boundary condition breaks parity (see Supplementary Note 2). In what follows, we will consider the generic case in which there is no degeneracy.

\begin{figure}
\centering
\includegraphics[width=\columnwidth]{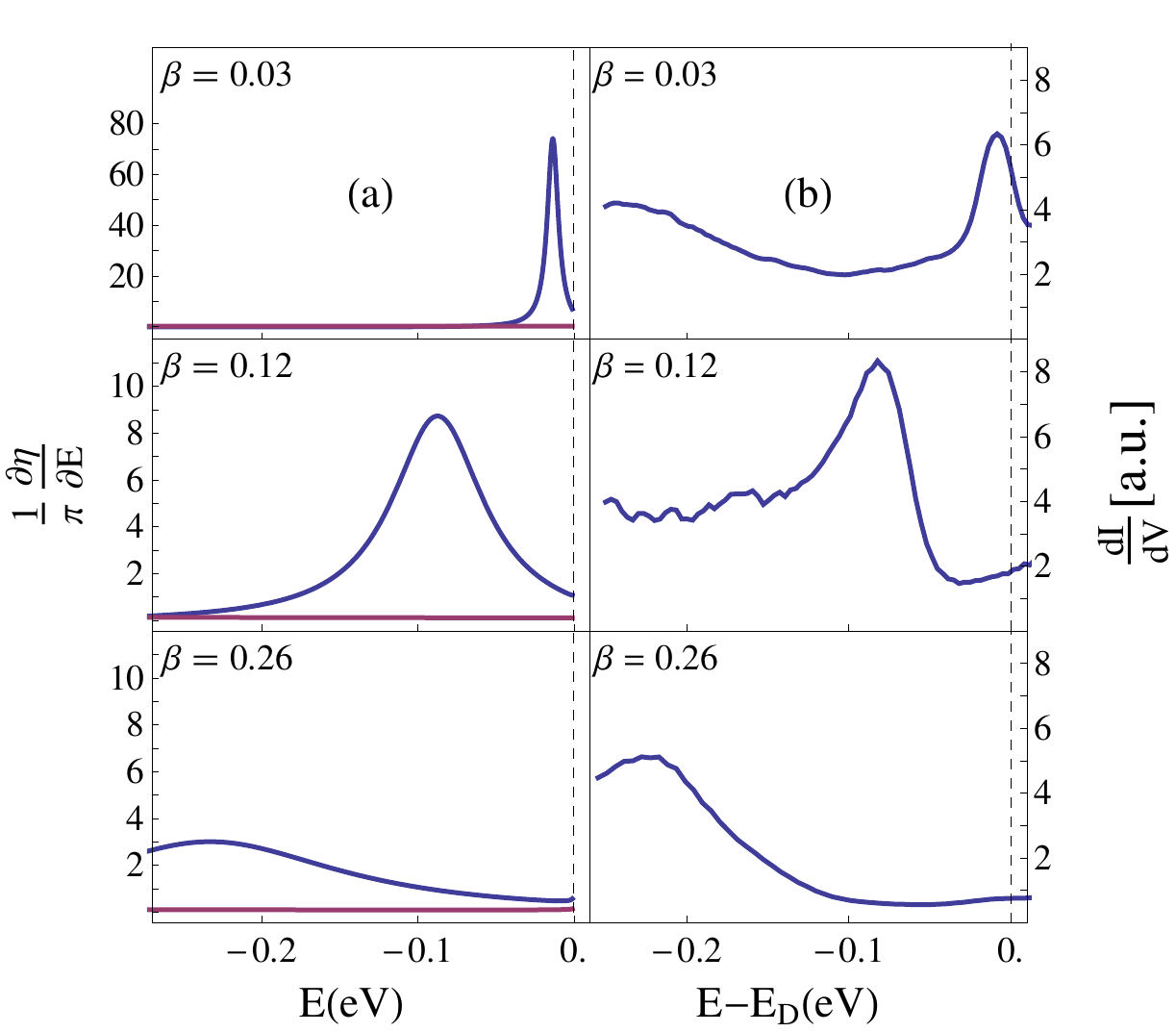}
\protect\caption{\label{Figqbs}
\textbf{Experimental and theoretical picture in the undercritical regime}. (a) Theoretical behaviour of $\frac{1}{\pi} d \eta/ d E$ for $d =2$ showing quasi-bound states of a massless Dirac fermion in the undercritical regime $\beta < 1/2$. In the scale-free low energy $E L \ll 1$ regime, the $m = -1$ (blue) branch contains a single peak and the $m = 0$ (purple) branch shows no peak independently of the choice of  boundary condition (see Supplementary Note 2). While increasing $\beta$, the resonance shifts to lower energy and becomes broader. (b) Excitation spectrum measured in graphene using STM as a function of the applied voltage $V$. The determination of the parameter $\beta$ is explained in the text. }
\end{figure}
\begin{figure}
\centering
\includegraphics[width=\columnwidth]{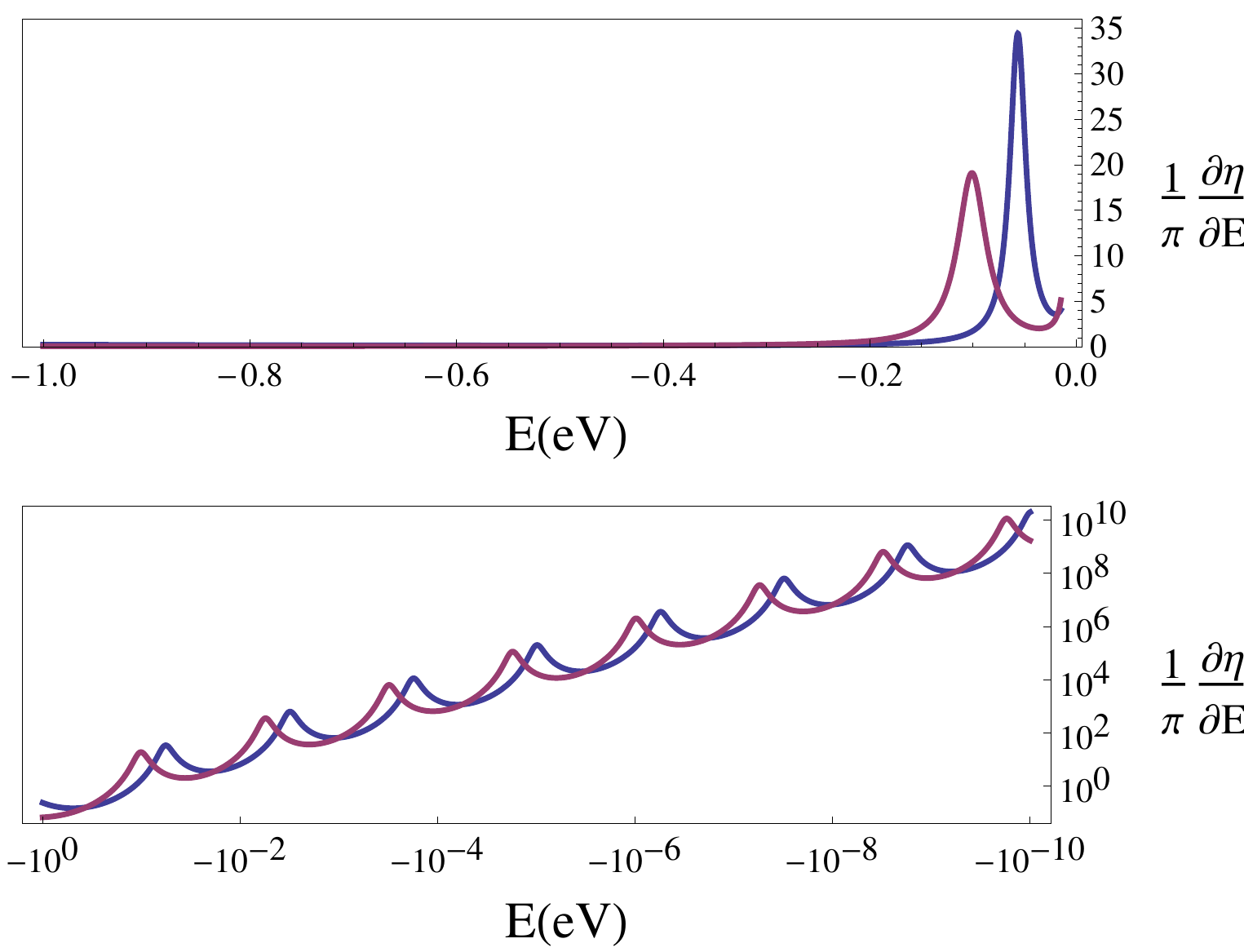}
\protect\caption{\label{FigqbsOverCrit}
\textbf{Theoretical behaviour of the low energy and scale free part of the quasi-bound states spectrum in the overcritical regime for $\boldsymbol{d = 2}$ and $\boldsymbol{\beta = 1.2 > \beta_{\mathrm{c}} (= 1/2) }$.} The lower plot displays the detailed structure of the infinite geometric ladders. Note that the $m = -1$ (blue) and $m = 0$ (purple) ladders are intertwined. These results are independent of the  boundary condition.}
\end{figure}

In the undercritical, $\beta  < \beta_{\mathrm{c}} $, and low energy regime $E L \ll 1$, we observe (see Figs.~\ref{fig1}b and \ref{Figqbs}a) a single quasi-bound state originating from only one of the $s$-wave channels and which broadens as $\beta$ increases.
In the overcritical regime $\beta > \beta_{\mathrm{c}}$, this picture changes dramatically.\footnote{We emphasize that this picture remains valid for all values of $\beta > \beta_{\mathrm{c}}$ and not only in the vicinity of $\beta_{\mathrm{c}}$} The low energy ($E L \ll 1$) scattering phase shift displays two intertwined, infinite geometric ladders of quasi-bound states (Figs.~\ref{fig1}c and \ref{FigqbsOverCrit}) at energies $E_n$ still given by (\ref{eq2}) but with $\zeta - \zeta_{\mathrm{c}}$ now replaced by $ \beta^2 - \beta_{\mathrm{c}} ^2 $. \footnote{Moreover, note that the energy scale $\epsilon_0$ for the Dirac case is different from the inverse square Schr\"odinger case defined in \eqref{eq1}.} This sharp transition at $\beta_{\mathrm{c}}$ belongs to the same universality class as presented for the inverse square Schr\"odinger problem, namely CS of the quasi-bound states spectrum is broken for $\beta > \beta_{\mathrm{c}}$ into a DS  phase characterized by a fractal distribution of quasi-bound states. The QPT thus reflects the lack of self-adjointness of the Hamiltonian (\ref{dirac}) and the necessary regularisation procedure leads to a scale anomaly in which CS is broken into DS.

\textbf{Experimental realization in graphene.} A particularly interesting condensed matter system where the previous considerations seem to be relevant is graphene in the presence of implanted Coulomb charges in conveniently created vacancies~\cite{mao2016realization}. It is indeed known that low energy excitations in graphene behave as a massless Dirac fermion field  with a linear dispersion $\epsilon = \pm v_F |p|$ and a Fermi velocity $v_F \simeq 10^6 \, \mathrm{m/s}$~\cite{EvaRev}. These characteristics have been extensively exploited to make graphene a very useful platform to emulate specific features of quantum field theory, topology and especially QED~\cite{shytov2007atomic}, since an effective fine structure constant $\alpha_G \equiv e^2 / \hbar v_F$ of order unity is obtained by replacing the velocity of light $c$ by $v_F$. 

It has been recently shown that single-atom vacancies in graphene can stably host local charge  \cite{mao2016realization}. Density functional theory (DFT) calculations have shown that when a carbon atom is removed from the honeycomb lattice, the atoms around the vacancy site rearrange into a lower energy configuration \cite{liu2014determining}. The resulting lattice reconstruction causes a charge redistribution which in the ground state has an effective local charge of $\approx +1$. Recent Kelvin probe force microscopy measurements of the local charge at the vacancy sites are in good agreement with the DFT predictions. Vacancies are generated by sputtering graphene with $\mathrm{He}^{+}$  ions~\cite{lehtinen2010effects,chen2011tunable}. Charge is modified and measured at the vacancy site by means of scanning tunnelling spectroscopy and Landau level spectroscopy as detailed in ~\cite{mao2016realization}. Applying multiple pulses allows for a gradual increase in the vacancy charge, which in turn acts as an effective tunable Coulomb source. Moreover, the size of the source inside the vacancy is small ($\approx 1\,\mathrm{nm}$) as compared to the method of deposited metal clusters~\cite{wang2013observing}. Using this method, we are able to observe the transition expected to occur at $\beta = 1/2$ and to measure and analyse three resonances for a broad range of $\beta$ values.

To establish a relation between the measured differential conductance and the spectrum of quasi-bound states, we recall that the tunnel current $I (V)$ is proportional to both  the density of states $\rho_t (\epsilon )$ of the STM tip and $\rho (\epsilon)$ of massless electronic excitations in graphene at the vacancy location. We  also assume that the tunnel matrix element $|t|^2$ depends only weakly on energy and that both voltage and temperature are small compared to the Fermi energy and height of the tunnelling potential, so that the current $I(V) = G_{\mathrm{t}} V$ is linear with $V$ thus defining the tunnel conductance $G_{\mathrm{t}} = 2 \pi \left( e^2 / \hbar \right) |t|^2 \rho_t \, \rho (\epsilon)$. Assuming that $\rho_t$ of the reference electrode (the tip) is energy independent, a variation $\delta \rho (\epsilon)$ of the local density of states at the vacancy leads to a variation $\delta I(V)$ of the current and thus to a variation $\delta G_{\mathrm{t}} (V)$ of the tunnel conductance so that, at zero temperature, we obtain~\cite{akkermans2007mesoscopic}
\be
 \frac{\delta G_\mathrm{t} (V)  }{ G_\mathrm{t}} = \frac{\delta \rho (\epsilon) }{ \rho_0} \, ,
 \label{tunnel} 
\ee
 where $\rho_0$  is the density of states in the absence of vacancy. 
 By considering the vacancy as a local perturbation, each quasi-particle state is  characterised by its scattering phase shift taken to be the phase shift $\eta (E )$ of the quasi-bound Dirac states previously calculated. Then, the change of density of states $\delta \rho (E )$ is obtained from (\ref{ks}) and  combining together with (\ref{tunnel}) leads to the relation, 
\be
\frac{d \delta I }{ dV } = \frac{G_{\mathrm{t}} }{ \pi \rho_0 } \frac{ d \eta (E ) }{ d E} 
\label{resonance}
\ee
between the differential tunnel conductance and the scattering phase shift. 
\begin{figure}
\centering
\includegraphics[width=\columnwidth]{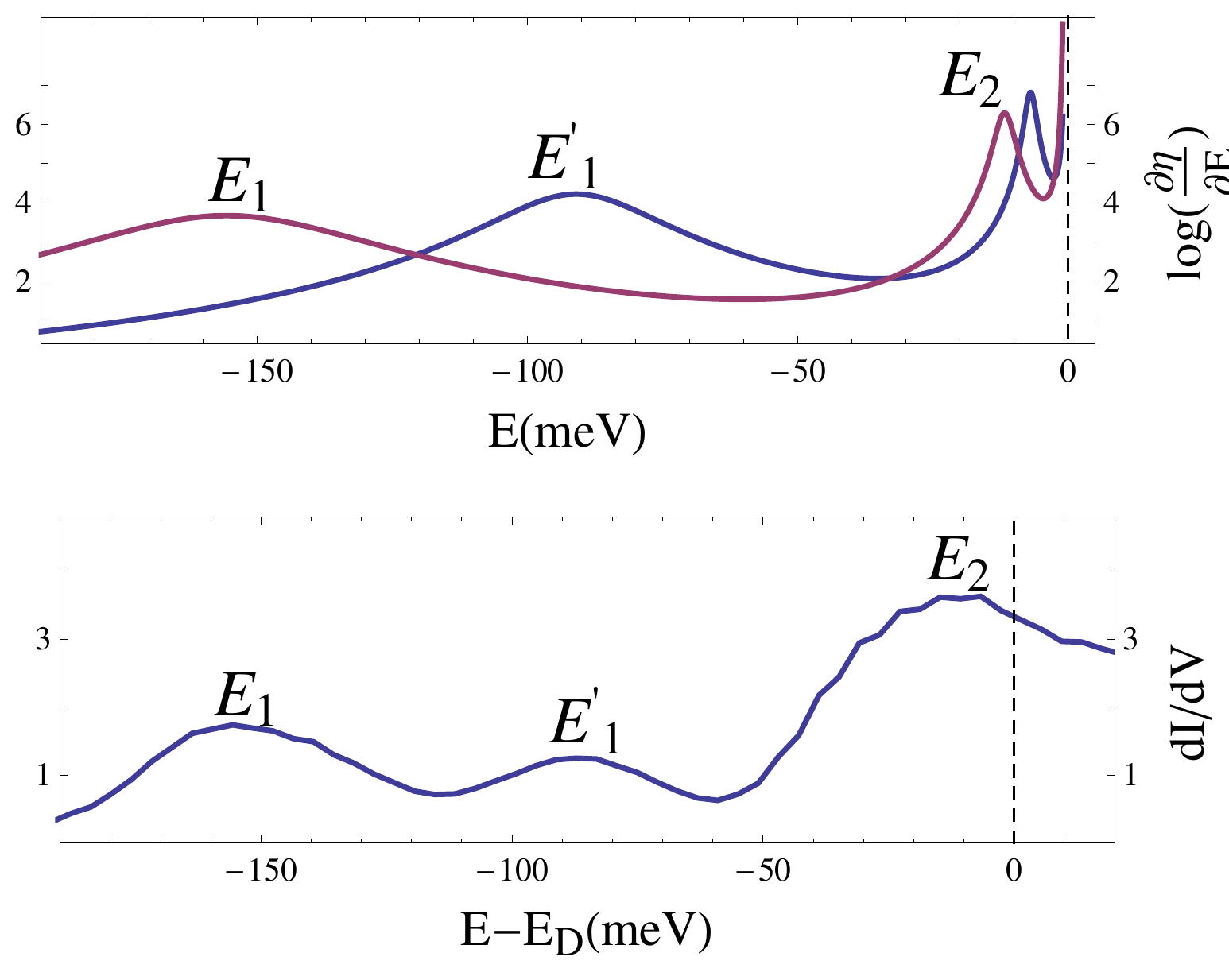}
\protect\caption{\label{Fig5}
\textbf{Experimental and theoretical picture in the overcritical regime.} Upper plot: Theoretical behaviour of the low energy and scale free part of the overcritical $(\beta = 1.33 )$ quasi-bound states spectrum obtained from (\ref{ks}). The blue (purple) line corresponds to $m = -1$ ($m = 0$). Lower plot: Experimental values of the (STM) tunnelling conductance measured at the position of charged vacancies in graphene. The labelling $E_1 , E_2 , E' _1$ of the peaks is explained in the text.
}
\end{figure}

\begin{figure}[h]
\includegraphics[width=\columnwidth]{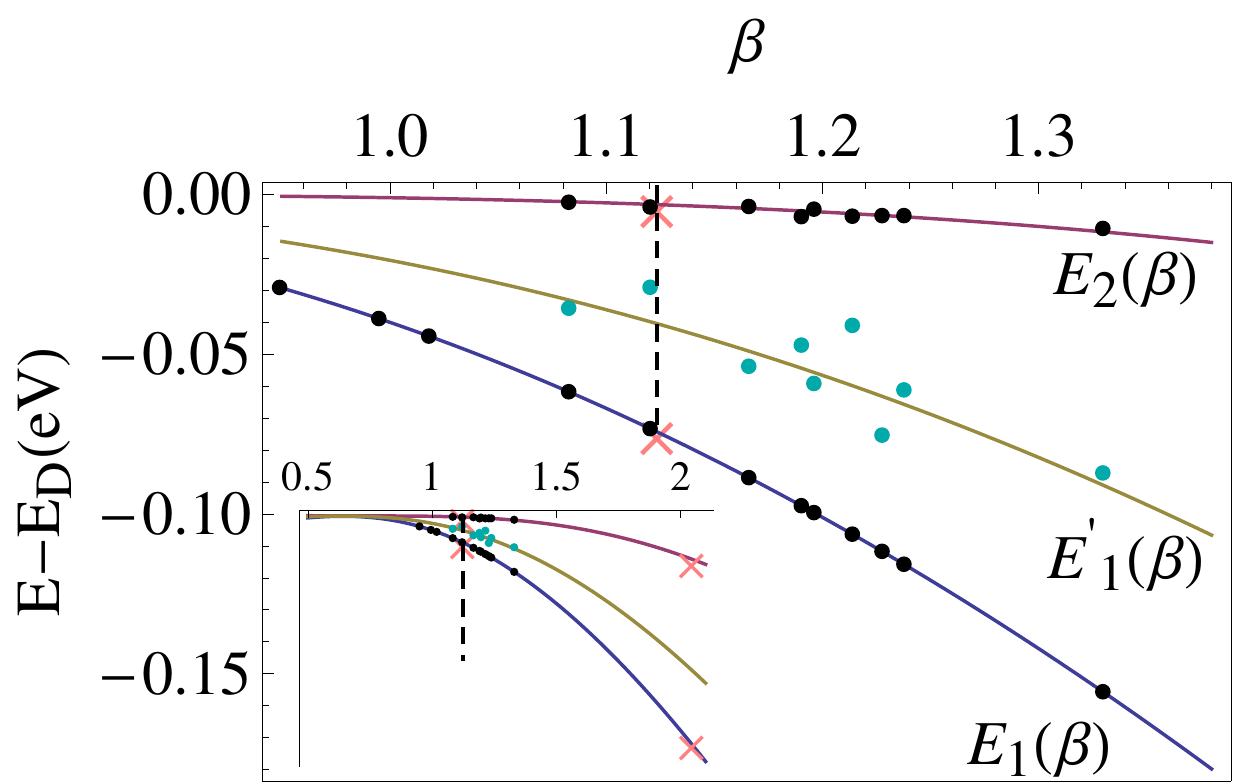}
\protect\caption{\label{Fig4}
\textbf{Behaviour of the energies $\boldsymbol{E_n (\beta)} $ of the quasi-bound state spectrum.} The curves are obtained from (\ref{eq2}) for $E_1(\beta), E'_{1}(\beta), E_2(\beta)$ as adapted to the massless Dirac case. The black and cyan dots correspond to the values measured in graphene. The two pink x's are the values of Efimov energies measured in Caesium atoms~\cite{kraemer2006evidence,huang2014observation} which corresponds to the (overcritical) fixed Efimov value $\beta_\mathrm{E} = 1.1236 $. Additional experimental points obtained in~\cite{tung2014geometric, Pires} are displayed in the inset.}
\end{figure}

\begin{figure}[h]
\centering
\includegraphics[width=\columnwidth]{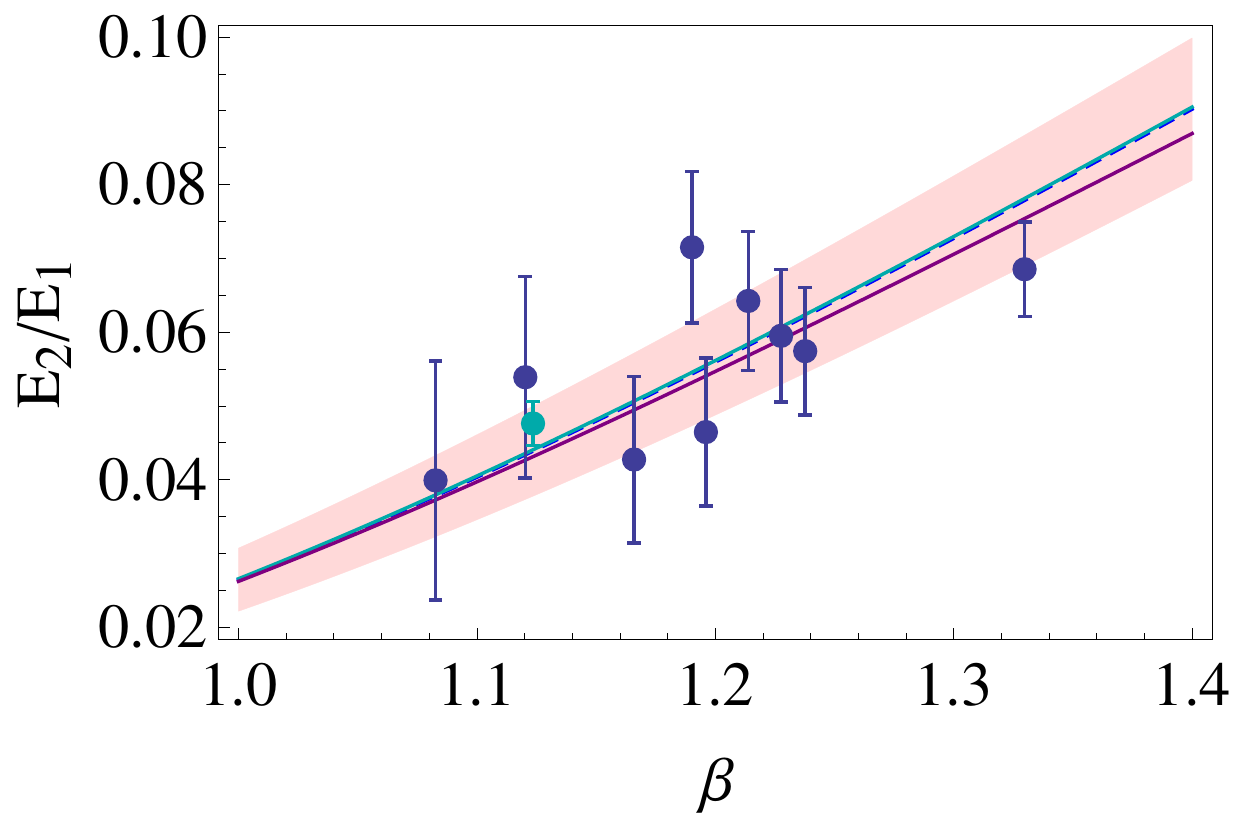}
\protect\caption{\label{theoryVsExperimentNew_Quantative} 
\textbf{Comparison between the experimentally obtained $E_2/E_1$ ratio and the universal factor $e^{-\pi / \sqrt{\beta^2 - 1/4}}$}.
Blue points: the ratio $E_2/E_1$ obtained from the position of the points in Fig.~\ref{Fig4}. Cyan point: Universal Efimov energy ratio as measured in Caesium atoms~\cite{kraemer2006evidence,huang2014observation}. Blue line (dashed): the corresponding optimized curve, fitted according to the model $e^{-b/\sqrt{\beta^2-1/4}}$ and corresponding to $b = 3.145$ with standard error of $\Delta b = 0.06$ consistent with the predicted value $\pi$. The shaded pink region is the $\pm 2 \Delta b$ confidence interval of the curve. Cyan line: universal low energy factor $e^{-\pi/\sqrt{\beta^2 - 1/4}}$. Purple line: theoretical ratio $E_2/E_1$ obtained from the exact solution of the Dirac equation. As $\beta \rightarrow 0.5 $, $|E_n|$ becomes smaller therefore the green and purple curves coincide for low $\beta$. The error bar on the resonance energies is $\pm 1mV$.}
\end{figure}

The measurements and data analysis presented here were carried out as follows:  positive charges are gradually injected into an initially prepared single atom vacancy and  the differential conductance $\delta G_{\mathrm{t}} (V)$ is measured at each step as a function of voltage. Since we are looking at the positions of resonant quasi-bound states, both quantities displayed in Figs.~\ref{Figqbs} and~\ref{Fig5} give the same set of resonant energies, independently of the energy-independent factor $G_{\mathrm{t}} / \pi \rho_0 $. 
For low enough values of the charge, the differential conductance displayed in Fig.~\ref{Figqbs}b, shows the existence of a single quasi-bound state resonance. The behaviour close to the Dirac point, namely in the low energy regime independent on the short distance regularization, is very similar to the theoretical prediction of Fig.~\ref{Figqbs}a. When the build up charge exceeds a certain value, we note the appearance of three resonances, emerging out of the Dirac point.  We interpret these resonances as the lowest overcritical $( \beta > 1/2 )$ resonances which we denote  $E_1, E'_{1}, E_2$ respectively. The corresponding theoretical and experimental behaviours displayed in Figs.~\ref{FigqbsOverCrit},~\ref{Fig5}, show a very good qualitative agreement. To achieve a quantitative comparison solely based on the previous Dirac Hamiltonian (\ref{dirac}), we fix $L$ and the boundary condition $h$ and deduce the theoretical $\beta$ values corresponding to the respective positions of the lowest overcritical resonance $E_1$ (as demonstrated in Fig.~\ref{Fig5}). This allows to determine the lowest branch $E_1 (\beta)$ for $n=1$ represented in Fig.~\ref{Fig4}.
Then, the experimental points $E'_{1}, E_2$ are directly compared to their corresponding theoretical branch as seen in Fig.~\ref{Fig4}. We determine $L$ and $h$, according to the ansatz $h = a(m + 1)$, and obtain the best correspondence for $L \simeq 0.2\,\mathrm{nm}$, $a \simeq -0.85$. We compare the experimental $E_2/E_1$ ratio with the universal prediction $E_{n+1} / E_n = e^{- \pi / \sqrt{\beta^2 - 1/4}}$ as seen in Fig.~\ref{theoryVsExperimentNew_Quantative}. A trend-line of the form $e^{-b/\sqrt{\beta^2-1/4}}$ is fitted to the ratios $E_2/E_1$ yielding a statistical value of $b = 3.145$ with standard error of $\Delta b = 0.06$ consistent with the predicted value $\pi$. An error of $\pm 1mV$ is assumed for the position of the energy resonances

A few comments are appropriate:
\begin{enumerate*}[label=(\roman*)]
\item \label{list:BC1} The points on the $E_2 (\beta)$ curve follow very closely the theoretical prediction $E_{n+1} / E_n = e^{- \pi / \sqrt{\beta^2 - 1/4}}$. This result is  insensitive to the choice of $h$, thus manifesting the universality of the ratio $E_{n+1} / E_n$.
\item \label{list:BC2} In contrast, the correspondence between the $E'_{1}$ points and the theoretical branch is sensitive to the choice of $h$. This reflects the fact that while each geometric ladder is of the form \eqref{eq2} (with the appropriate $\zeta \rightarrow \beta$ change), the energy scale $\epsilon_0$ is different between the two thus leading to a shifted relative position of the two geometric ladders in Fig.~\ref{FigqbsOverCrit}. The ansatz taken for $h$ is phenomenological (see Supplementary Note 2), however, we find that in order to get reasonable correspondence to theory, the explicit dependence on $m$ is needed. More importantly, it is necessary to use a degeneracy breaking boundary condition to describe the $E'_{1}(\beta)$ points. For instance, if the Coulomb potential is regularised by a constant potential for $r \leq L$~\cite{pereira2008supercritical}, then both angular momentum channels (i.e., the $E'_{1}$ and $E_1$ points) become degenerate. The existence of the experimental $E'_{1}$ branch is therefore a distinct signal that parity symmetry in the corresponding Dirac description (\ref{dirac}) is broken. In graphene, exchanging the triangular sublattices is equivalent to a parity transformation. Creating a vacancy breaks the symmetry between the two sub-lattices and is therefore at the origin of broken parity in the Dirac model.
\item \label{list:BC3} The value $L \simeq 0.2 \, \mathrm{nm}$ is fully consistent with the low energy requirement $E_1 L / \hbar v_F \simeq 0.03 \ll 1$ necessary to be in the regime relevant to observe the $\beta$-driven QPT.
\end{enumerate*}

\subsection*{Discussion}
A further argument in support of the universality of this QPT is achieved by comparing the experimental results obtained in graphene with those deduced from a completely different physical problem. To that purpose, we dwell for a short while recalling the basics underlying Efimov physics~\cite{Braaten2006259}. Back to 1970, Efimov~\cite{efimov1970energy} studied the quantum problem of three identical nucleons of mass $m$ interacting through a short range $(r_0)$ potential. He pointed out that when the scattering length $a$ of the two-body interaction becomes very large, $a \gg r_0$, there exists a scale free regime for the low energy spectrum, $\hbar^2 / m a^2 \ll E \ll \hbar^2 / m r_0 ^2 $, where the corresponding bound states energies follow the geometric series ($\sqrt{-E_n} = - \tilde \epsilon_0 e^{- \pi n / s_0}$) where $s_0 \simeq 1.00624$ is a dimensionless number and $\tilde \epsilon_0$ a problem-dependent energy scale. Efimov deduced these results from an effective Schr\"odinger equation in $d=3$ with the radial $(l=0)$ attractive potential $V(r) = - \left( s_0 ^2 + 1/4 \right) / r^2$. Using equations \eqref{eq1}, \eqref{eq2} and the critical value $\zeta_{\mathrm{c}} = ( d-2 )^2 / 4 = 1/4$ for this Schr\"odinger problem, we deduce the $\zeta$  value for the Efimov effect to be $s_0 ^2 + 1/4 > \zeta_{\mathrm{c}}$ corresponding to the overcritical regime of the QPT. The value of $\beta$ matching to the Efimov geometric series factor $e^{ \pi / s_0}$ is $\beta_\mathrm{E} = \sqrt{s_0^2+1/4} = 1.1236$, referred to as the fixed Efimov value. Despite being initially controversial, Efimov physics  has turned into an active field especially in atomic and molecular physics where the universal spectrum has been studied experimentally~\cite{kraemer2006evidence,pollack2009universality,Gross2009,Lompe2010a,Nakajima2011,tung2014geometric,Pires,Kunitski2015} and theoretically~\cite{Braaten2006259}. The first two Efimov states $E_n$ $( n=1 , 2)$ have been recently determined using an ultracold gas of caesium atoms~\cite{huang2014observation}. Although the Efimov spectrum always lies at a fixed and overcritical value of the coupling, unlike the case of graphene where $\beta$ can be tuned,  the universal character of the overcritical regime allows nevertheless for a direct comparison of these two extremely remote physical systems. To that purpose, we include the Efimov value $\beta_\mathrm{E}$ in the expression obtained for the massless Dirac fermion in a Coulomb potential and insert the corresponding data points obtained for cold atomic caesium in the graphene plot (Fig.~\ref{Fig4}) up to an appropriate scaling of $\tilde{\epsilon}_0$. The results are fully consistent thus showing in another way the universality presented. 

There are other remote examples of systems displaying this universal QPT, e.g., flavoured QED3~\cite{appelquist1988critical}, and the XY model (Kosterlitz-Thouless~\cite{Kaplan:2009kr} and roughening transitions~\cite{PhysRevB.46.12664}). Our results provide a useful and original probe of  characteristic features of this universal QPT and motivate a more thorough study of this transition.

\section*{Methods}

Our sample is stacked two layers of graphene on top of a thin BN flake (see Fig.~\ref{fig1}f). The standard dry transfer procedure is followed to get this heterostructure. A large twisted angle between the two layers graphene is selected in order to weaken the coupling. The free-standing like feature for the top layer graphene is checked by the Landau levels spectroscopy. To achieve the diluted single vacancies, the sample is exposed to the helium ion beam for short time (100\,eV for 5\,s) followed by the high temperature annealing. The experiment is performed at 4.2\,K with a home-built STM. The $dI/dV$ ($I$ is the current, $V$ is the bias) is recorded by the standard lock-in technique, with a small AC modulation 2\,mV at 473.1\,Hz added on the DC bias. To tune the effective charge on the vacancy, we apply the voltage pulse (-2\,V, 100\,ms) with the STM tip directly locating on top of the vacancy.

\subparagraph*{Data availability} The data that support the findings of this study are available from the corresponding author upon request.

\section*{Acknowledgments} 
This work was supported by the Israel Science Foundation Grant No. 924/09. Funding for the experimental work  provided by DOE-FG02-99ER45742 (STM/STS), NSF DMR 1207108 (fabrication and characterization).

\section*{Author contributions} 
O.~Ovdat and E.~Akkermans proposed observing the aforementioned quantum phase transition in graphene with a charged vacancy. They have contributed to interpreting and solving the theoretical model as well as analysing the experimental data and making contact with the theory. J.~Mao, Y.~Jiang and E.~Y.~Andrei conceived of and designed the experiment, as well as performed the measurements and analysed the data. All authors contributed to discussions and preparation of the
manuscript.

\subparagraph*{Competing Financial Interests} 
The authors declare no competing financial interests.


\bibliography{references}

\clearpage


\section*{Supplementary material}

We present in supplementary figures 1 and 2 all overcritical and undercritical measurements, and their corresponding theoretical plots. In supplementary figure 3 we present the STM topography image of an isolated vacancy in graphene.

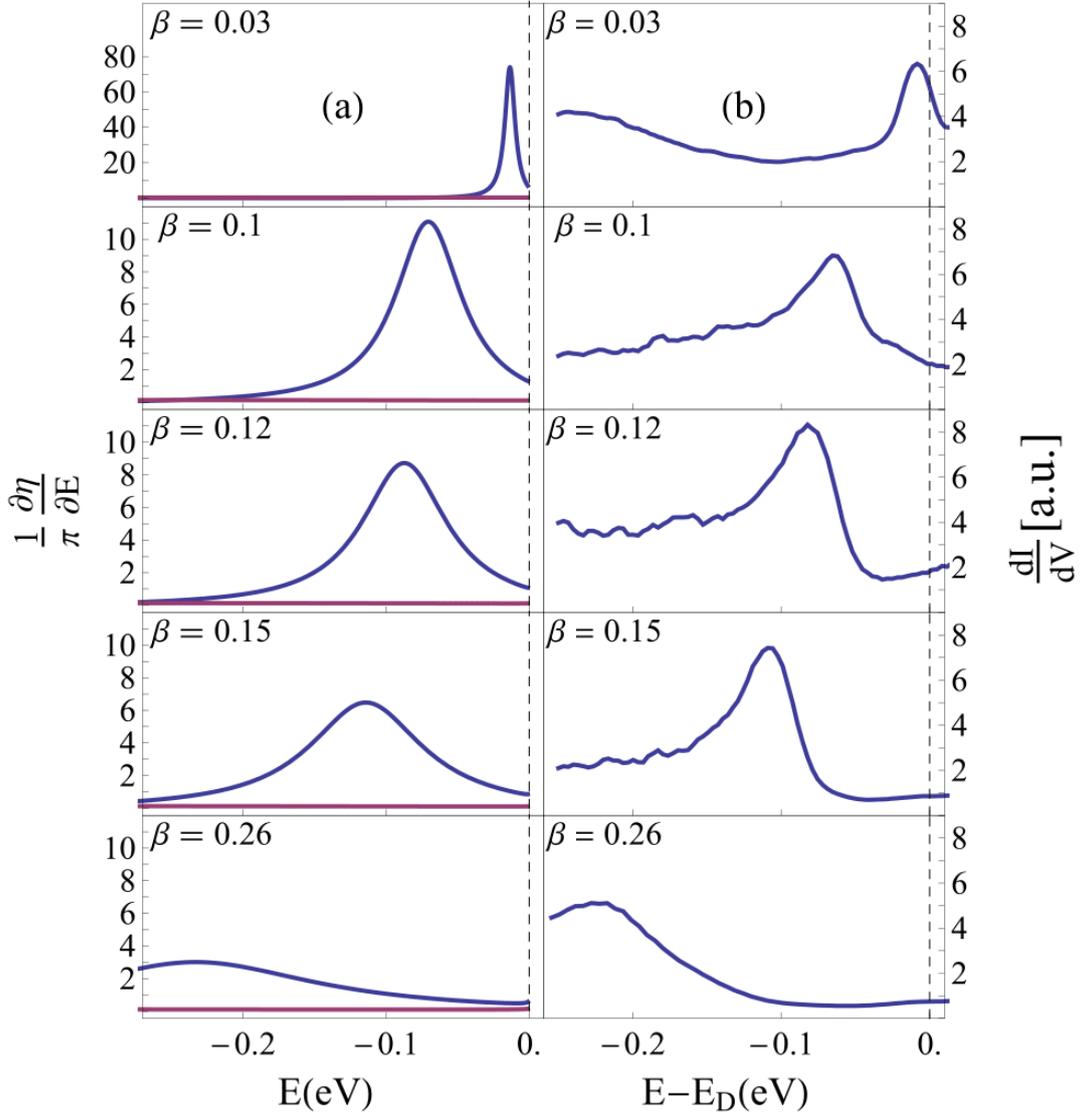

Supplementary Figure 1: **Experimental and theoretical picture in the undercritical regime.** a. Theoretical behaviour of the undercritical ($\beta < 1/2$) quasi-bound states spectrum of massless Dirac fermions. b. STM measurement of the quasi-particle spectrum at the position of the charged vacancy in the undercritical regime. See main text for more details.

## Supplementary Note 1: Dirac Coulomb problem in $d+1$ dimensions – critical coupling, phase shift and quasi bound states

In what follows, we study a system described by a massless Dirac particle in the presence of an electric potential that has an inverse radial tail in $d + 1$ dimensions. We show that beyond a critical coupling value, an anomalous breaking of conformal symmetry occurs and the system is described by a discrete scale invariant spectrum. We obtain for a general over critical coupling and short range behaviour of the potential, an expression for an infinite series of geometrically spaced quasi bound states.

The massless Dirac equation with an attractive potential $V(r) = -\frac{\beta}{r}$, $\beta \equiv Z\alpha$ in $d+1$ dimensions is

$$i\gamma^\mu \left(\partial_\mu + ieA_\mu\right)\psi\left(x^\nu\right) = 0 \tag{1}$$

where $\mu = 0 \ldots d$, $A_\mu$ is the electromagnetic potential (EM)

$$\begin{aligned} eA_0 &= -\beta/r \\ A_i &= 0 \quad i = 1, \ldots, d. \end{aligned} \tag{2}$$

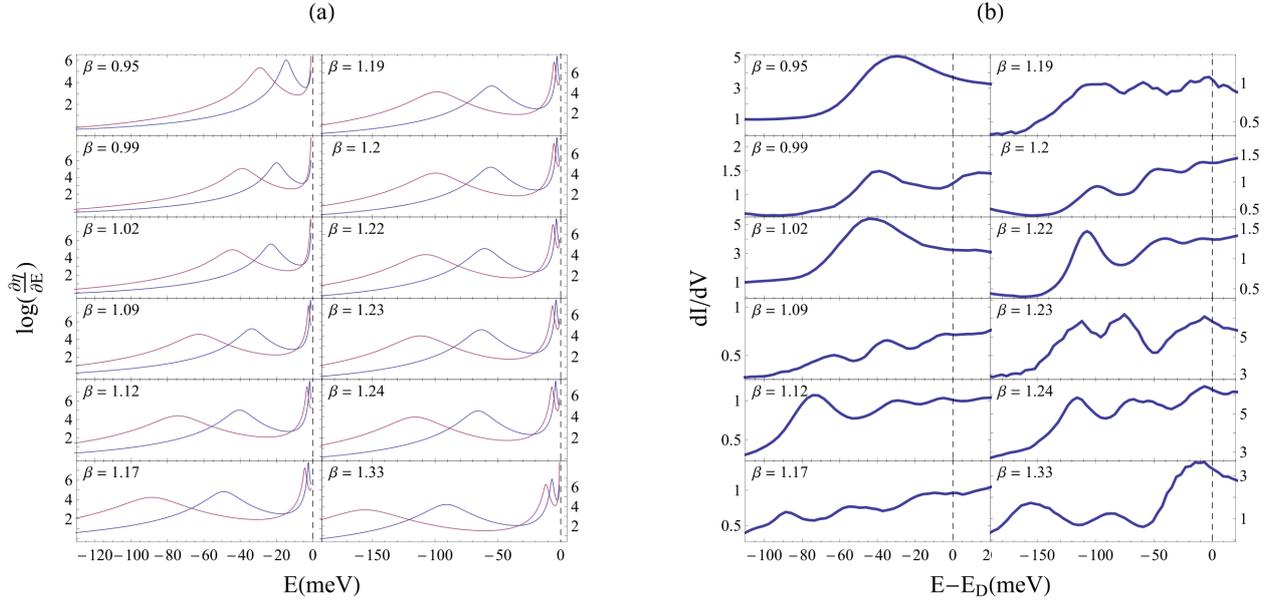

Supplementary Figure 2: **Experimental and theoretical picture in the overcritical regime.** a. Theoretical behaviour of the overcritical ($\beta > 1/2$) quasi-bound states spectrum of massless Dirac fermions. b. STM measurement of the quasiparticle spectrum at the position of the charged vacancy in the overcritical regime. See main text for more details.

and $\gamma^\mu$ are $d+1$ matrices satisfying the anti-commutation relation

$$\{\gamma^\mu, \gamma^\nu\} = 2\eta^{\mu\nu} \tag{3}$$

with $\eta_{\mu\nu}$ being the $d+1$ Minkowski metric with a 'mostly minus' sign convention. The Hamiltonian of the system is expressed as

$$H = \gamma^0 \gamma^j p_j - \beta/r \tag{4}$$

where $j = 1\ldots d$ and corresponds to the scale invariant eigenvalue equation $H\psi = E\psi$ equivalent to (1).

Utilizing rotational symmetry, the angular part of supplementary equation (1) can be solved and the radial dependence of $\psi(x^\nu)$ is given in terms of two functions $\Psi_2(r), \Psi_1(r)$ [1] determined by the following set of equations

$$\Psi_2'(r) + \frac{(d-1+2K)}{2r}\Psi_2(r) = \left(E + \frac{\beta}{r}\right)\Psi_1$$
$$-\Psi_1'(r) - \frac{(d-1-2K)}{2r}\Psi_1(r) = \left(E + \frac{\beta}{r}\right)\Psi_2 \tag{5}$$

where

$$K \equiv \begin{cases} \pm\left(l + \frac{d-1}{2}\right) & d > 2 \\ m + 1/2 & d = 2 \end{cases}, \tag{6}$$

$l = 0, 1, \ldots$ and $m \in \mathbb{Z}$ are orbital angular momentum quantum numbers. In terms of these radial functions, the scalar product of two eigenfunctions $\psi, \tilde\psi$ is given by

$$\int dV\, \psi^\dagger \tilde\psi = \int dr\, r^{d-1} \left(\Psi_1^*(r)\tilde\Psi_1(r) + \Psi_2^*(r)\tilde\Psi_2(r)\right).$$

We introduce a short distance radial cut-off $L$ and assume that there exist an electric Coulomb potential $V(r) = -\frac{\beta}{r}$ for $r > L$ and some interaction at $r < L$ that can be modelled by a BC at $r = L$. The equivalent mixed boundary condition of (5) can be written as follows [2]

$$h = \left.\frac{\Psi_2(r)}{\Psi_1(r)}\right|_{r\to L^+} \tag{7}$$

where $h$ is determined by the short range physics and in general can depend on $E, L$ and $K$. Specification of the cut-off $L$, coupling $\beta$, boundary condition $h$ (and the already determined angular dependence) determine a specific set of solutions to equation (1).

Two independent solutions to equations (5) are given by $S_{\pm\gamma}$ with

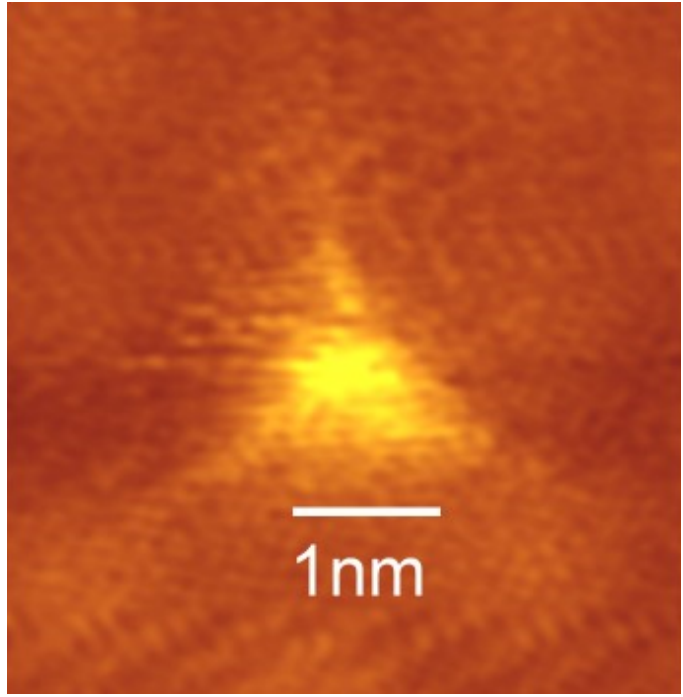

Supplementary Figure 3: **Characteristic topography signature of an isolated vacancy in graphene.** The triangular interference pattern arises due to the local crystal distortion and corresponding electronic state reconstruction. This is the feature that was used to identify single atom vacancies in this work.

$$S_\gamma(\rho) \equiv r^{\frac{1-d}{2}} e^{-\rho/2} \rho^{i\gamma} \left[ {}_1F_1\left(i(\gamma+\beta); 1+2i\gamma; \rho\right) \begin{pmatrix} 1 \\ i \end{pmatrix} \right.$$
$$\left. + \frac{\gamma+\beta}{K} {}_1F_1\left(1+i(\gamma+\beta); 1+2i\gamma; \rho\right) \begin{pmatrix} i \\ 1 \end{pmatrix} \right] \quad (8)$$

where ${}_1F_1(a,b,z)$ is Kummer's function [3], $\gamma \equiv \sqrt{\beta^2 - K^2}$ and $\rho \equiv 2iEr$. The $|E|r \ll 1$ behaviour of $S_{\pm\gamma}$ is given by

$$S_\gamma = r^{\frac{1-d}{2}} \rho^{i\gamma} \left( \begin{pmatrix} 1+i\frac{\gamma+\beta}{K} \\ i+\frac{\gamma+\beta}{K} \end{pmatrix} + \mathcal{O}(|E|r) \right). \quad (9)$$

Solutions corresponding to outgoing and ingoing radial waves for $r \to \infty$ are given by the combinations

$$\psi_{\text{in}} = \left(C_{\text{in}}^\gamma S_\gamma + C_{\text{in}}^{-\gamma} S_{-\gamma}\right)$$
$$= r^{\frac{1-d}{2}} \left( \begin{pmatrix} 1 \\ -i \end{pmatrix} e^{-iEr - i\beta \log(2|E|r)} + \mathcal{O}\left(\frac{1}{|E|r}\right) \right)$$
$$\psi_{\text{sc}} = \left(C_{\text{sc}}^\gamma S_\gamma + C_{\text{sc}}^{-\gamma} S_{-\gamma}\right)$$
$$= r^{\frac{1-d}{2}} \left( \begin{pmatrix} 1 \\ i \end{pmatrix} e^{+iEr + i\beta \log(2|E|r)} + \mathcal{O}\left(\frac{1}{|E|r}\right) \right) \quad (10)$$

where

$$C_{\text{sc}}^\gamma \equiv K e^{-\frac{\pi\beta}{2}} \frac{\left(e^{2\pi\beta} - e^{-2\pi\gamma}\right) \Gamma(i(\gamma+\beta))}{\left(e^{2\pi\gamma} - e^{-2\pi\gamma}\right) \Gamma(1+2i\gamma)}$$
$$C_{\text{in}}^\gamma \equiv e^{\pi\gamma} e^{\frac{3\pi\beta}{2}} \frac{\left(e^{-2\pi\beta} - e^{-2\pi\gamma}\right) \Gamma(1+i(\gamma-\beta))}{\left(e^{2\pi\gamma} - e^{-2\pi\gamma}\right) \Gamma(1+2i\gamma)} \quad (11)$$

For the case $E < 0$. The log dependence in (10) is characteristic of the long range Coulomb tail and is irrelevant to the physics of the scattering problem [4]. Thus, a general solution to (5) can be written as

$$\begin{pmatrix} \Psi_1 \\ \Psi_2 \end{pmatrix} \propto \psi_{\text{in}}(\rho) + e^{2i\eta} \psi_{\text{sc}}(\rho) \quad (12)$$

where the scattering phase shift $\eta$ is determined by the boundary condition (7) at $r = L$

$$e^{2i\eta(EL,h)} = -\left.\frac{\psi_{in,2}(\rho) - h\psi_{in,1}(\rho)}{\psi_{sc,2}(\rho) - h\psi_{sc,1}(\rho)}\right|_{\rho=2iEL} \quad (13)$$

The energy derivative of $\eta$ is given by the logarithmic derivative of the RHS of (13)

$$\frac{d\eta}{dE} = \frac{1}{2i}\frac{d}{dE}\ln\left(-\frac{\psi_{in,2}(\rho) - h\psi_{in,1}(\rho)}{\psi_{sc,2}(\rho) - h\psi_{sc,1}(\rho)}\bigg|_{\rho=2iEL}\right) \quad (14)$$

The explicit expression is complicated and is therefore omitted. Resonant quasi-bound states appear for specific values of the energy $E$ at which $d\eta/dE$ exhibits a sharp maxima. We would like to obtain an analytic expression for the position of the resonances in the regime $|E|L \ll 1$ appearing in Fig. 3 of the main text. Such an expression can in principle be extracted from (14). However, an easier route is available by allowing the energy parameter to be complex valued such that $E \to \varepsilon \equiv E_R - i\frac{W}{2}$ [5] and look for solutions of (5) and (7) with no $e^{-iEr}$ plane wave solution for $r \to \infty$. The lifetime of the resonance is given by $W^{-1}$ and is required to be positive. A solution with no ingoing wave is obtained by the requirement that $e^{-2i\eta(EL,h)} = 0$ or alternatively the vanishing of the denominator in (13) such that

$$h = \frac{\psi_{sc,2}(\rho)}{\psi_{sc,1}(\rho)}\bigg|_{\rho=2i\varepsilon L}. \quad (15)$$

Focusing our attention at the regime $|\varepsilon|L \ll 1$, one finds very different results depending on the value of $\beta$. For $\beta < \beta_c \equiv |K|$, $\gamma$ is pure imaginary. Thus, from (9) and (10) we get that both $\psi_{sc,1,2} \propto \rho^{-\sqrt{K^2-\beta^2}}$ and therefore equation (15) is independent of $\varepsilon$ to leading order in $|\varepsilon|L$. This inconsistency means that for fixed $L$ and $\beta$ there are no quasi bound states arbitrarily close to zero energy. In contrary, for $\beta > \beta_c$ and to leading order in $|\varepsilon|L$, equation (15) is given by

$$h_0 = \frac{C_s^\gamma \rho^{i\gamma}\left(i + \frac{\beta+\gamma}{K}\right) + (\gamma \to -\gamma)}{C_s^\gamma \rho^{i\gamma}\left(1 + i\frac{\beta+\gamma}{K}\right) + (\gamma \to -\gamma)} \quad (16)$$

where $h_0 \equiv h|_{E\to 0}$. Solving supplementary equation (16) for $\rho^{2i\gamma}$ gives $\rho^{2i\gamma} = z_0$ where $z_0 \equiv \frac{C_s^{-\gamma}((1-ih_0)(\beta-\gamma)-(h_0-i)K)}{C_s^\gamma((h_0-i)K-(1-ih_0)(\beta+\gamma))}$. Inserting $\rho \equiv 2i\varepsilon L$ and solving for $\varepsilon$ yields

$$\varepsilon_n = -\epsilon_0 e^{i\Theta} e^{-\frac{\pi n}{\gamma}} \quad n \in \mathbb{Z} \quad (17)$$

where $\varepsilon_0 \equiv \frac{1}{2L}|z_0^{\frac{1}{2i\gamma}}| > 0$ and $\Theta \equiv \arg\left(iz_0^{\frac{1}{2i\gamma}}\right)$. The regime of validity of this result is for $n > n_{min}$, where $n_{min}$ is such that $|\varepsilon_{n_{min}}|L \sim 1$. Using $\Gamma$-function identities, the phase $\Theta$ of $-\varepsilon_n$ can be simplified to

$$\begin{aligned}\Theta &= \arg\left(iz_0^{\frac{1}{2i\gamma}}\right) \\ &= \frac{\pi}{2} - \frac{1}{2\gamma}\log(|z_0|) \\ &= -\frac{\pi}{2} + \frac{1}{4\gamma}\log\left(\frac{\sinh(\pi(\beta+\gamma))}{\sinh(\pi(\beta-\gamma))}\right).\end{aligned} \quad (18)$$

Note that $\Theta$ is independent on the boundary condition $h_0$. Recall that $\gamma \equiv \sqrt{\beta^2 - K^2}$, where $K$ is given in (6) for $d \geq 2$. For $\beta > |K| \geq \frac{1}{2}$, $\arg(-\varepsilon_n)$ is a very slowly varying function bounded in the region $0 < \arg(-\varepsilon_n) < 0.046\pi$, as can be seen in supplementary figure 4. This is consistent with negative resonances $E_n = \mathrm{Re}(\varepsilon_n) < 0$ with width and $W_n \equiv -2\mathrm{Im}(\varepsilon_n) > 0$. Therefore, in the regime $|\varepsilon|L \ll 1$, $\varepsilon_n$ describes a geometrically spaced set of infinitely many negative quasi bound states for all $\beta > |K|$, $h, d \geq 2$. The width of the resonance is given by $W_n \equiv -2\mathrm{Im}(\varepsilon_n) < 0.9\pi|\varepsilon_n|$ which describes resonances getting sharper indefinitely as $n \to \infty$. In the vicinity of the transition point $\beta - |K| \gtrsim 0$, $\arg(-\varepsilon_n) \cong -\frac{\pi}{1-e^{2|K|\pi}} + \mathcal{O}(\xi - |K|)^{1/2}$ which for $d = 2$, $m = 0 (\Leftrightarrow |K| = \frac{1}{2})$ agrees with the result of reference [6] in the main text.

## Supplamentary Note 2: Dirac Coulomb system as a model for a charged vacancy in graphene

As explained in the main text, we model excitations around the vacancy as massless Dirac particles and neglect any interactions between them. Therefore, for $r > L$, $L$ being the characteristic size of the vacancy, they are described by the Dirac equation (5) with $d = 2$. The full spinor in this case is given in polar coordinates by

$$\psi(r,\phi) = e^{im\phi}\begin{pmatrix}\Psi_1(r) \\ i\Psi_2(r)e^{i\phi}\end{pmatrix} \quad (19)$$

in a representation where $H = \vec{\sigma}\cdot\vec{p} - \beta/r$. The components of the spinor correspond to amplitudes of the wave function on each of the two graphene sublattices. In the near vacancy region, $r < L$ we assume some unknown interaction that can be taken into account by the boundary condition (7) at $r = L$. Conventionally used choices are:

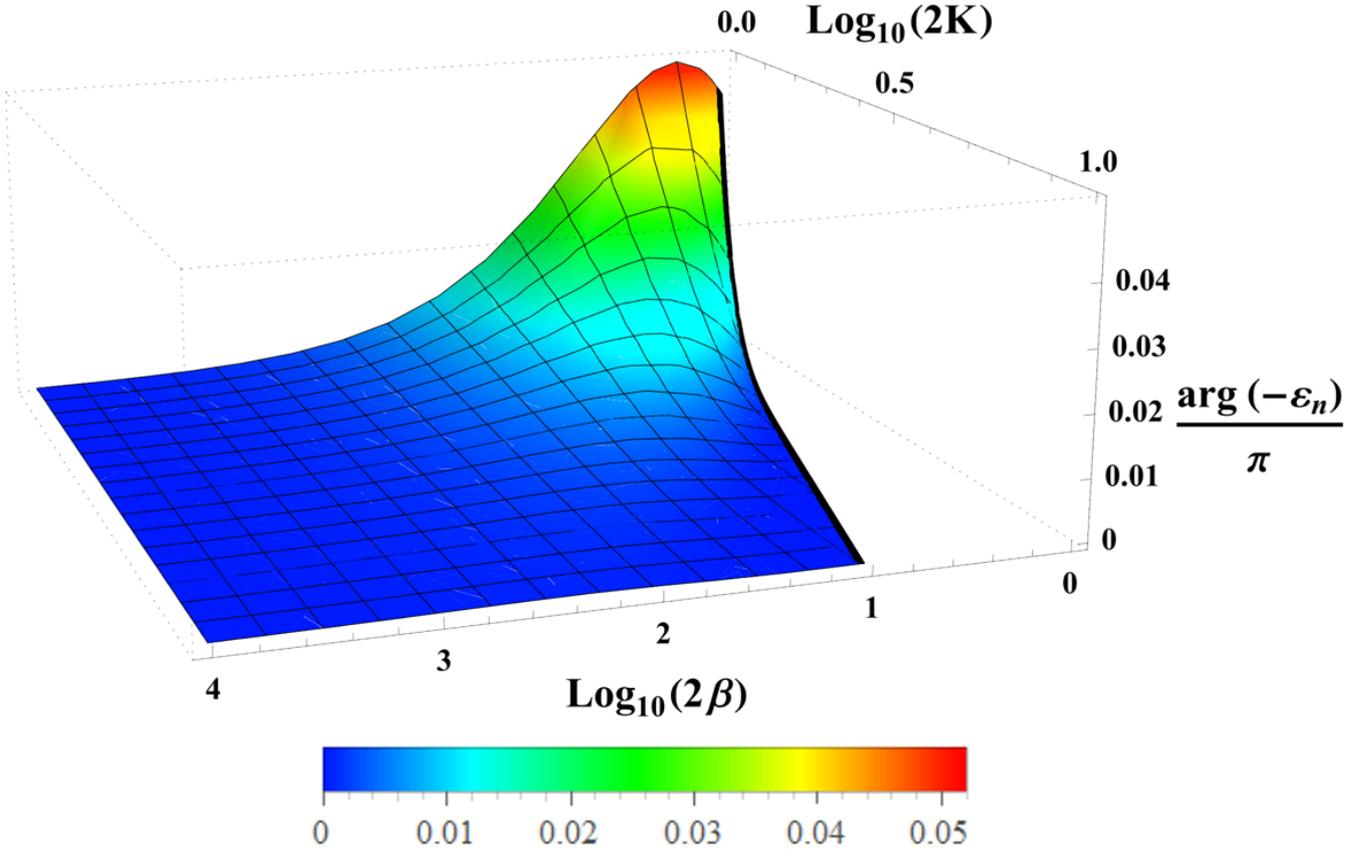

Supplementary Figure 4: **The phase of $-\varepsilon_n$ as a function of the coupling $\beta$ and $|K|$ for $\beta > |K|$.** The bold black line represents $\xi = |K|$. It can be seen that the phase of $-\varepsilon_n$ takes values in the range $[0, 0.046\pi]$ for $d \geq 2$.

(i) continuously connected constant potential $V_< = -\beta/L$ [7] corresponding to $h = J_{m+1}(\beta + EL)/J_m(\beta + EL)$, where $J_n(x)$ is Bessel's function; (ii) zero wavefunction on one of the graphene lattice sites [8] corresponding to $h = 0$; (iii) infinite mass term on boundary term [9] corresponding to $h = 1$. Generically, $h$ can depend on $E, L$ and $m$. From (6), the critical coupling is $\beta_c \equiv |m + 1/2| \leq 1/2$, giving rise to two angular momentum s-wave channels, $m = 0, -1$ for which $\beta_c = 1/2$.

An additional important property of our model is related to parity symmetry. Since mass or scalar potential terms are absent (and the Coulomb potential is radial), the Dirac equation in $2 + 1$ dimension is symmetric under 2 dimensional parity, $P_y$, in which $(x, y) \to (-x, y)$ [10]. The action of the parity operators on $\Psi(r, \phi)$ is defined as

$$\psi'(r, \phi) \equiv P_y \psi(r, \pi - \phi)$$
$$= \sigma_y \psi(r, \pi - \phi)$$
$$= -i e^{i(-m-1)\phi} \begin{pmatrix} \Psi_2(r) \\ -i\Psi_1(r)e^{i\phi} \end{pmatrix}. \quad (20)$$

This transformation can also be accounted for (up to an unimportant overall phase) by

$$\Psi_1(r) \to \Psi_2(r), \Psi_2(r) \to -\Psi_1(r), m \to -m - 1 \quad (21)$$

where $K \equiv m + 1/2$, thus $m \to -m - 1 \Leftrightarrow K \to -K$. Indeed, the Dirac equation (5), is invariant under (21).

From (21) it is apparent that solutions corresponding to angular momentum $m$ and $-m-1$ are linked via parity symmetry. Since energy remains unchanged under parity, a natural question is whether the quasi bound energies are the same for angular momentum channels $m$ and $-m-1$. The answer depends on the short distance region. For example, if we describe the $r < L$ regime by condition (i) then the quasi bound states will necessarily be degenerate over the $m$ and $-m-1$ channels. The reason is that in this case, the potential of both regimes $r > L, r < L$ respects parity symmetry. As a result, $h$ in (i) transforms like $\Psi_2(r)/\Psi_1(r)$ under (21), i.e., $h \to -h^{-1}$. Thus, by applying (21) on both side of (14) it is straightforward to obtain that $d\eta_m/dE = d\eta_{-m-1}/dE$. However, if the potential in the $r < L$ regime break parity, that is, the corresponding boundary condition does not transform as $h \to -h^{-1}$ under (21) (for example $h = 0, 1$), then $d\eta_m/dE \neq d\eta_{-m-1}/dE$ and the degeneracy will be broken. Specifically, for $m = 0, -1$ we find two interleaved geometric ladders of over critical states in the corresponding regime $\beta > \beta_c = 1/2$ as shown in the main text. The relative position of the two ladders typically depends on $h$ and therefore is sensitive to the detail of the short range physics.

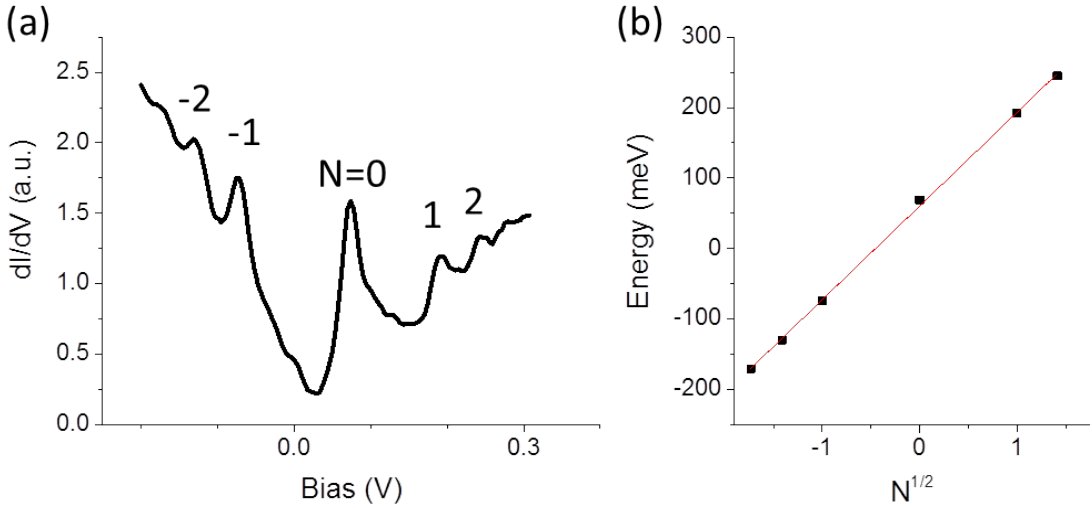

Supplementary Figure 5: **The Landau levels spectrum for twisted bilayer graphene under** 10T. (a) dI/dV curve on graphene at $B = 10$T with $V_b = -300$mV and $I = 20$pA. (b) Fit of the Landau level sequence in (a) used to extract the Fermi velocity.

All the figures in the main text describing the behaviour of $d\eta/dE$ as a function of $E$ and the values of the quasi bound states $E_n(\beta)$ are extracted from the exact relation appearing in equation (14).

## Supplementary methods

**Sample fabrication**. In this work, we use G/G/BN on $SiO_2$ to perform the experiment. The hBN thin flakes were exfoliated onto the $SiO_2$ surface followed by a dry transfer process using a sacrificial PMMA thin film to stack the first graphene layer on the hBN flake. Before stacking the top layer graphene, the PMMA was removed with acetone and IPA, followed by furnace annealing in forming gas (10% $H_2$ and 90% Ar) at 230°C for 3 hours. The second layer graphene was stacked by using the same procedure as the first layer. Au/Ti electrodes were deposited by the standard SEM lithography for the STM contact. After the liftoff process, the sample was annealed again in furnace with forming gas to remove the PMMA residues. Subsequently, the sample is loaded in the UHV chamber for further annealing at $230°C$ overnight. To generate the single vacancies, the sample is exposed to a $100 - 140$eV He$^+$ ion beam followed by high temperature *in situ* annealing. The other stacked samples, G/BN/$SiO_2$ and G/G/$SiO_2$, were prepared by the same procedure.

**Characterization by STM topography and Landau level (LL) spectroscopy**. The STM experiment is performed at 4.2K using a cut PtIr tip. The dI/dV spectroscopy is performed using the standard lockin method with bias modulation typically 2mV at 473.1Hz. To charge the single vacancies, voltage pulses are applied directly at the desired vacancy site with the STM tip at ground potential. The intrinsic electronic properties of graphene can be effectively isolated from the random potential induced by the $SiO_2$ substrate by using an intermediate graphene layer and an hBN buffer underneath the layers. When the twist angle between the two stacked graphene layers exceeds 10° the two layers are electronically decoupled at the experimentally relevant energies. Therefore a large twist angle was chosen to ensure a linear dispersion near the Dirac point. LL spectroscopy provides a direct way to prove the layer decoupling. For single layer graphene, the energy level sequence is given by: $E_N = \text{sign}(N)v_F\sqrt{2e\hbar|N|B}$, where $N$ is the LL index, $v_F$ is the Fermi velocity, $\hbar$ is the reduced Plank constant. For fixed magnetic field, the root N dependence for the sequence is the fingerprint of single layer graphene. supplementary figure 5(a) shows the LL spectrum of the G/G/BN sample at 10T. By fitting the LLs sequence (supplementary figure 5(b)), we obtain the value of the Fermi velocity $v_F = (1.2 \pm 0.02) \times 10^6 m/s$.

# Supplementary References





\end{document}